# Bell violation with entangled photons, free of the fair-sampling assumption†

**Authors:** Marissa Giustina[1,2]‡, Alexandra Mech[1,2]‡, Sven Ramelow[1,2]‡, Bernhard Wittmann[1,2]‡, Johannes Kofler[1,3], Jörn Beyer[4], Adriana Lita[5], Brice Calkins[5], Thomas Gerrits[5], Sae Woo Nam[5], Rupert Ursin[1], Anton Zeilinger*[1,2]

**Affiliations:**

[1]Institute for Quantum Optics and Quantum Information – Vienna (IQOQI), Austrian Academy of Sciences, Boltzmanngasse 3, Vienna, Austria.

[2]Quantum Optics, Quantum Nanophysics, Quantum Information, University of Vienna, Faculty of Physics, Boltzmanngasse 5, Vienna, Austria.

[3]Max Planck Institute of Quantum Optics (MPQ), Hans-Kopfermannstr. 1, 85748 Garching, Germany.

[4]Physikalisch-Technische Bundesanstalt, Abbestraße 1, 10587 Berlin, Germany.

[5]National Institute of Standards and Technology (NIST), 325 Broadway, Boulder, CO 80305, USA.

*Correspondence to: anton.zeilinger@univie.ac.at

† Partial contribution of NIST, an agency of the U.S. government, not subject to copyright.

‡These authors contributed equally to this work.

**Abstract**: The violation of a Bell inequality is an experimental observation that forces one to abandon a local realistic worldview, namely, one in which physical properties are (probabilistically) defined prior to and independent of measurement and no physical influence can propagate faster than the speed of light. All such experimental violations require additional assumptions depending on their specific construction making them vulnerable to so-called "loopholes." Here, we use photons and high-efficiency superconducting detectors to violate a Bell inequality closing the fair-sampling loophole, i.e. without assuming that the sample of measured photons accurately represents the entire ensemble. Additionally, we demonstrate that our setup can realize one-sided device-independent quantum key distribution on both sides. This represents a significant advance relevant to both fundamental tests and promising quantum applications.

**Introduction:** In 1935, Einstein, Podolsky, and Rosen (EPR) (*1*) argued that quantum mechanics is incomplete when assuming that no physical influence can be faster than the speed of light and that properties of physical systems are elements of reality. They considered measurements on spatially separated pairs of entangled particles. Measurement on one particle of an entangled pair projects the other instantly on a well-defined state, independent of their spatial separation. In 1964, Bell (*2*) showed that no local realistic theory can reproduce all quantum mechanical predictions for entangled states. His renowned Bell inequality proved that there is an upper limit to the strength of the observed correlations predicted by local realistic theories. Quantum theory's predictions violate this limit.

In a Bell experiment, one prepares pairs of entangled particles and sends them to two observers, Alice and Bob, for measurement and detection. Alice and Bob observe correlations between their results that, for specific choices of their measurement settings, violate the Bell inequality and therefore force abandonment of local realism.

It is common that in an experiment, some particles emitted by the source will not be detected (*3, 4*). Then the subset of detected particles might display correlations that violate the Bell inequality although the entire ensemble can be described by a local realistic theory. To achieve a conclusive Bell violation without assuming that the detected particles are a "fair" sample, a highly efficient experimental setup is necessary. As noted by Pearle in (*3*), this efficiency need not be perfect.

Due to experimental limitations, fair sampling has been assumed in nearly every Bell experiment performed to date; a few exceptions include (*5-8*). To date, it has never been possible to avoid this assumption with photons due to the absence of efficient sources and detectors. Here we report the first Bell experiment with photons that does not rely on any fair-sampling assumption.

Since the first experimental Bell test (*9*), a satisfactory laboratory realization of the motivating gedankenexperiment has remained an outstanding challenge. The two other main assumptions include "locality" (*10, 11*) and "freedom of choice" (*12*). Invoking any of these renders an experiment vulnerable to explanation by a local realistic theory. The realization of an experiment that is free of all three assumptions – a so-called loophole-free Bell test – remains an important outstanding goal for the physics community (*13*). We note that with our experiment, photons are the first physical system for which each of these three assumptions has been successfully addressed, albeit in different experiments.

In our experiment we employ the Eberhard inequality, a Bell inequality which by construction does not rely on the fair-sampling assumption (*14*). Our source of photon pairs utilizes spontaneous parametric down-conversion in a Sagnac configuration; this construction has proved to be efficient (*15, 16*). For photon detection, we employ superconducting transition-edge sensors (TES) which not only offer high efficiency but also are intrinsically free of dark counts (*17*). This combination of features is imperative for an experiment in which no correction of count rates can be tolerated.

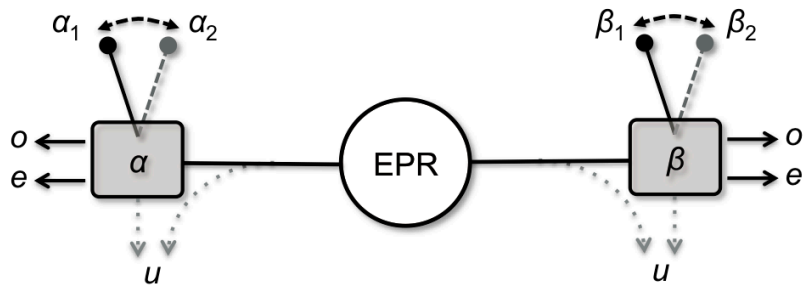

**Fig. 1:** Principle of the experiment. Violation of an Eberhard inequality involves an EPR (Einstein, Podolsky, Rosen) source of polarization-entangled pairs as well as polarization measurements. Each measurement device can rotate the photon's polarization according to one of two settings ($\alpha_1$, $\alpha_2$ and $\beta_1$, $\beta_2$) before projecting the photon in the "ordinary" (*o*) or

"extraordinary" (*e*) output of a polarizing beam splitter and detecting it. All lost photons are also included in the derivation of the inequality as "undetected" (*u*) events. The terms of the inequality are photon counts recorded in the different settings.

**Theory:** Eberhard's inequality, which was proposed almost two decades ago (*14*), is a CH-type Bell-inequality (*18*) that explicitly includes also undetected (inconclusive) events. Therefore, its violation already implies that the fair-sampling loophole is closed. Moreover, the derivation of Eberhard's inequality even includes pairs not detected on either side (and can be generalized for those not even produced), which means that no post-selection on the created pairs is necessary to violate the inequality.

It is broadly recognized that Eberhard's inequality requires the lowest known symmetric arm efficiency for non-maximally entangled qubit states of only $\eta = 2/3 \approx 66.7$ %. This arm efficiency comprises all losses, in particular those in the source and the measurement setup including the detector. Note that with asymmetric efficiencies or higher-dimensionally entangled states, thresholds lower than 2/3 have been reported (*19*, *20*). For the most widely used Bell inequality by Clauser, Horne, Shimony and Holt (CHSH) (*21*) at least $\eta = 2\sqrt{2}-2 \approx 82.8$ % is necessary in the symmetric case. For polarization-entangled photon-pairs, Eberhard's inequality considers three possible outcomes: *o* ("ordinary") and *e* ("extraordinary") for the two recorded outcomes of a polarization measurement, and *u* ("undetected") if no photon is detected (see fig. 1). Two different measurement settings $\alpha_1$, $\alpha_2$ ($\beta_1$, $\beta_2$) on Alice's (Bob's) side are used, respectively. Let $n_{kl}(\alpha_i,\beta_j)$ denote the number of pairs with the outcome $k$ ($l$) for Alice's (Bob's) photon with $k,l \in \{o,e,u\}$, when measured in settings $\alpha_i$ ($\beta_j$) with $i,j \in \{1,2\}$. Then the Eberhard inequality can be written as:

$$J = -n_{oo}(\alpha_1,\beta_1) + n_{oe}(\alpha_1,\beta_2) + n_{ou}(\alpha_1,\beta_2) + n_{eo}(\alpha_2,\beta_1) + n_{uo}(\alpha_2,\beta_1) + n_{oo}(\alpha_2,\beta_2) \geq 0 \qquad [1]$$

Local realism allows $J$ to take only non-negative values. Quantum-mechanically, the maximal violation is given by $J/N = (1-\sqrt{2})/2 \approx -0.207$ (*22*), where $N$ denotes the number of entangled particle pairs produced per applied setting combination. This bound is reachable for $\eta = 1$ symmetric arm efficiency and maximally entangled states. For the largest possible violation of the Eberhard inequality with $\eta < 1$, non-maximally entangled states must be used. These take the form

$$|\psi_r\rangle = \frac{1}{\sqrt{1+r^2}}\left(|HV\rangle + r|VH\rangle\right) \qquad [2]$$

with $0 < r < 1$ and $H$ ($V$) denoting horizontal (vertical) polarization of Alice's and Bob's photons. Depending on the background count rate, efficiencies higher than $\eta = 2/3$ are needed (*14*). Interestingly, for $\eta < 82.8$ %, non-maximally entangled states are not only optimal but even necessary for a violation of Eberhard's inequality.

In an experiment, one records measurements of "singles counts" $S$ (number of detection events on one side) and coincidence counts $C$ (number of detected pairs) for the four different combinations of settings $(\alpha_1,\beta_1)$, $(\alpha_1,\beta_2)$, $(\alpha_2,\beta_1)$, and $(\alpha_2,\beta_2)$. The number of events where one of the outcomes is undetected follows directly from the measured rates: for a given measurement length, let us denote the measured coincidence counts by $C_{kl}(\alpha_i,\beta_j)$ and the singles counts by $S_k^A(\alpha_i)$ for Alice and $S_l^B(\beta_j)$ for Bob ($k,l \in \{o,e\}$). Then all terms in Eberhard's inequality are given by the measured quantities as follows:

$n_{oo}(\alpha_1,\beta_1) = C_{oo}(\alpha_1,\beta_1)$

$n_{oe}(\alpha_1,\beta_2) = C_{oe}(\alpha_1,\beta_2)$  $\quad n_{ou}(\alpha_1,\beta_2) = S_o^A(\alpha_1) - C_{oo}(\alpha_1,\beta_2) - C_{oe}(\alpha_1,\beta_2)$

$n_{eo}(\alpha_2,\beta_1) = C_{eo}(\alpha_2,\beta_1)$  $\quad n_{uo}(\alpha_2,\beta_1) = S_o^B(\beta_1) - C_{oo}(\alpha_2,\beta_1) - C_{eo}(\alpha_2,\beta_1)$

$n_{oo}(\alpha_2,\beta_2) = C_{oo}(\alpha_2,\beta_2)$

Inserting this into Eberhard's inequality yields:

$$J = -C_{oo}(\alpha_1,\beta_1) + S_o^A(\alpha_1) - C_{oo}(\alpha_1,\beta_2) + S_o^B(\beta_1) - C_{oo}(\alpha_2,\beta_1) + C_{oo}(\alpha_2,\beta_2) \geq 0 \qquad [3]$$

where the coincidence counts $C_{oe}(\alpha_1,\beta_2)$ and $C_{eo}(\alpha_2,\beta_1)$ have dropped out. The resulting inequality that is used in our experiment now only contains directly available detection events related to the ordinary beams of Alice and Bob. Remarkably, this implies that Alice and Bob each only need one detector to test the inequality while closing the fair-sampling loophole, in contrast to two detectors each for testing a CHSH inequality. This feature can also be intuitively understood: consider detectors that also monitor the *e* outcomes, and gradually lower their detection efficiency until it vanishes completely. This will just move events from "*e*" to "*u*", i.e. from $n_{oe}(\alpha_1,\beta_2)$ to $n_{ou}(\alpha_1,\beta_2)$ and from $n_{eo}(\alpha_2,\beta_1)$ to $n_{uo}(\alpha_2,\beta_1)$. Since only their sum appears in the Eberhard inequality, the value of *J* does not change.

**Experiment:** The entangled photon pairs at 810 nm are produced in a Sagnac type source (*15, 23*) pumped by a 405 nm laser. The source is based on type-II spontaneous parametric down-conversion using a nonlinear crystal (ppKTP). In each arm, a cut-off filter and a 3 nm interference filter with near 99 % transmission are used to suppress counts from the pump laser and reduce the background counts. The source can be tuned to produce non-maximally entangled states of the form [2] for any *r* by setting the polarization of the pump light with a half- and quarter-wave plate.

The measurement setup (see fig. 2), containing a rotatable half-wave plate in a high-precision rotation mount and a calcite polarizer, is stationed in front of the fiber coupler on both Alice´s and Bob´s side to facilitate measurement of the desired polarization (*α* and *β*). Only one output of the polarizer is needed for the measurements (see fig. 2 and theory), therefore only the transmitted ordinary beam of the polarizer is coupled into the fiber. The extraordinary beam is blocked after transmission. We couple the 810 nm photons into an optical fiber (SMF-28), which guides the photons to the sensitive area of the detectors. To achieve high coupling efficiency in both arms, we optimized the focusing of the pump laser and the fiber couplers (*24, 25*).

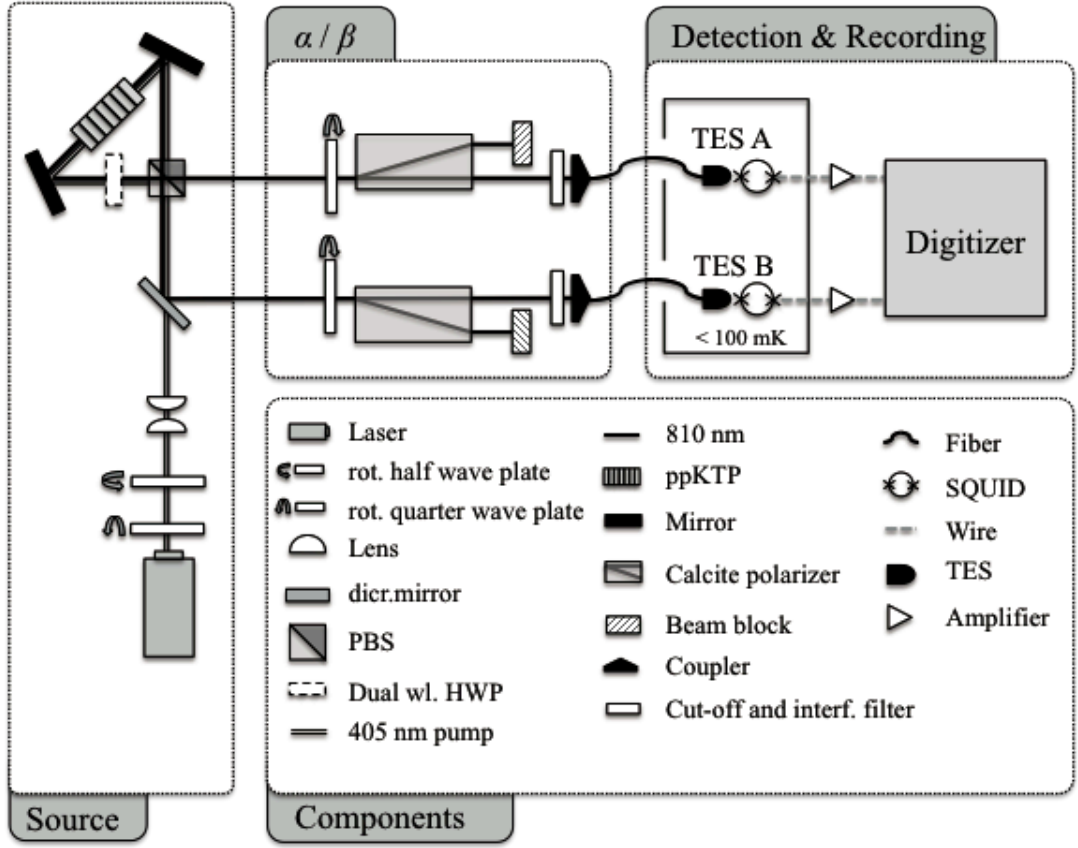

**Fig. 2:** Setup. The source, based on spontaneous parametric down-conversion in a Sagnac-configuration, produces polarization-entangled photons at 810 nm. A measurement setting is implemented in each arm by rotating a half-wave plate to the desired angle in front of a calcite polarizer. Photons transmitted through the calcite polarizer (ordinary output beam) are filtered spectrally and coupled into an optical fiber (SMF-28) which leads them to transition-edge sensors (TES) for detection. The output signals from the detectors are amplified by superconducting quantum interference devices (SQUIDs) and further electronics before being digitized and processed by an algorithm that identifies photons and time-correlated photon pairs.

To achieve highly efficient photon detection, we employed TES calorimetric detectors that owe their sensitivity to operation at the superconducting transition, a regime characterized by steep R-T dependence (*17*). Benefiting from a wavelength-optimized optical structure, these detectors have been reported to demonstrate detection efficiencies of up to 98%, including losses from packaging and fiber coupling (*16, 17*). Superconducting quantum interference devices (SQUIDs) (*26*) amplify the nA-scale TES current signal, which is subsequently digitized and stored for later analysis. Algorithms identify photon signatures in the analog output signal, determine an arrival time for each event, and count two-photon coincidences, all without requiring additional information from the user about the data.

As a guide for the experimental settings needed to observe a violation of local realism, we used numerical simulations and optimization to determine an optical non-maximally entangled state. For input, the model used overall efficiencies $\eta_A$ and $\eta_B$ at Alice and Bob's side, the estimated background rate, and the observed visibility. The model not only estimated a value for $r$ but also appropriate measurement settings $\alpha_1$, $\alpha_2$, $\beta_1$, and $\beta_2$ at Alice and Bob's side.

**Result:** We used a value of ~0.3 for $r$ and measured for a total of 300 seconds per setting at each of the four settings $\alpha_1\beta_1$, $\alpha_1\beta_2$, $\alpha_2\beta_1$, and $\alpha_2\beta_2$ described by angles $\alpha_1 = 85.6°$, $\alpha_2 = 118.0°$, $\beta_1 = -5.4°$, and $\beta_2 = 25.9°$. The relevant singles and coincidence counts obtained appear below and yield a $J$-value of $J = -126715$.

| $C_{oo}(\alpha_1,\beta_1)$ | $S_o^A(\alpha_1)$ | $C_{oo}(\alpha_1,\beta_2)$ | $S_o^B(\beta_1)$ | $C_{oo}(\alpha_2,\beta_1)$ | $C_{oo}(\alpha_2,\beta_2)$ | $J$ |
|---|---|---|---|---|---|---|
| 1069306 | 1522865 | 1152595 | 1693718 | 1191146 | 69749 | –126715 |

**Table 1:** Measurement results and $J$-value for a total measurement time of 300 s per setting. Without background subtraction, the Eberhard $J$-value can be calculated from the measured data according to [3]. Green (red) values contribute beneficially (detrimentally) to a negative $J$-value.

After recording for a total of 300 seconds per setting we divided our data into 10-second blocks and calculated the standard deviation of the resulting 30 different $J$-values. This yields a sigma of 1837 for our aggregate $J$-value of $J = -126715$, a 69-$\sigma$ violation. Note that this calculation does not assume Poissonian counting statistics or any error propagation rules. We estimate the number of produced pairs to $N = 24.2 \cdot 10^6$ per applied setting, yielding a normalized violation of $J/N = -0.00524$ ($\pm$ 0.00008).

Under the assumptions of locality and freedom-of-choice, a negative $J$-value refutes local realism without the fair-sampling assumption or post-selection on created pairs, regardless of the states and angles used for the measurement or any error in their implementation. Nonetheless, additional measurements can provide further insight to the obtained value. The directly measured arm efficiencies (each a ratio of observed coincidence and singles counts without any correction) measured in the $HV$-basis were $\eta_A = 73.77\%$ ($\pm$ 0.07 %) in Alice's arm and $\eta_B = 78.59\%$ ($\pm$ 0.08 %) in Bob's. We attribute imperfect coupling efficiency to a variety of possibly arm-dependent effects including optical losses in the source, coupling, fiber splices, and detectors. We estimate our $r$-value and visibility to around 0.297 and 97.5 % respectively. Using these values, our numerical model (used for the aforementioned optimization) agrees very well with our measured $J$-value.

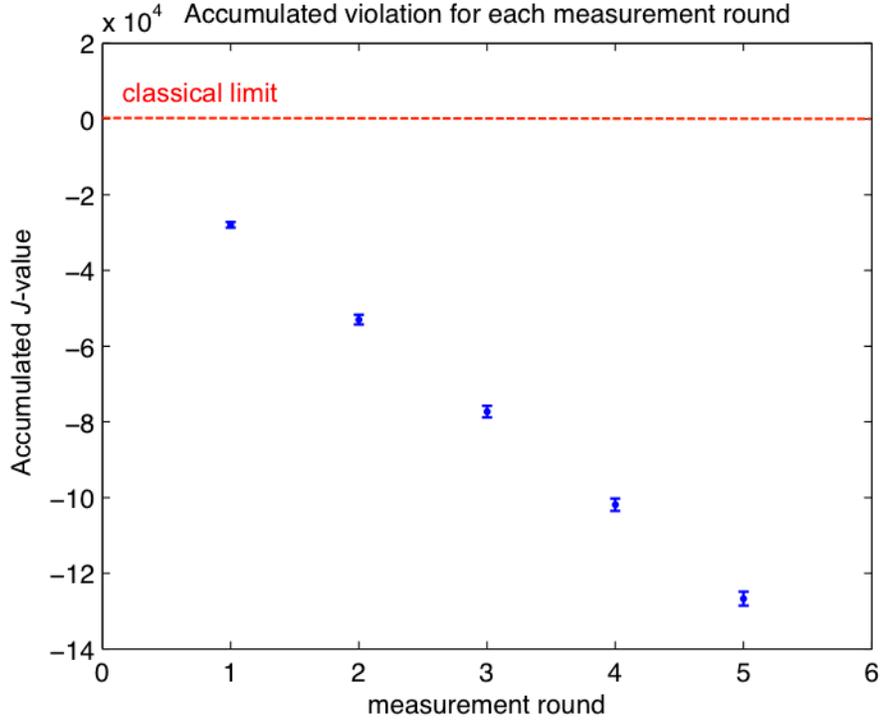

**Fig. 3:** Eberhard *J*-value computed from up to five measurements of recorded data. Any negative *J*-value violates the inequality and refutes all local realistic models that exploit the fair-sampling loophole. Error bars represent plus/minus one standard deviation calculated from the binned raw data.

Closing the fair-sampling loophole is not only important for answering fundamental questions about science but is also relevant to practical applications like device-independent quantum key distribution (DI-QKD) (*27*). The basic idea in DI-QKD protocols is that they do not require assumptions about the operation of any physical apparatus. The practical relevance of this has already been demonstrated, as photon detectors can be manipulated by an outsider even to mimic experimental Bell-violations based on explicitly exploiting the fair-sampling loophole (*28*). DI-QKD is inherently immune to such attacks. To implement DI-QKD, at least 75 % arm efficiency is needed (*29*). Our results demonstrate that this bound is within the reach of current technology.

With an additional assumption, namely that a communicating party trusts his own measurement apparatus, one arrives at one-sided DI-QKD where the arm efficiency requirement is reduced to 65.9 % (*30*). This protocol enables a communication party to convince himself that he shares entangled photons with his partner, even if the partner and source are not trusted. Implementing this protocol on both sides allows both partners the same peace of mind, as *each* can confirm their security relying only on devices within their own control.

Additional measurements demonstrate that our apparatus is capable of implementing one-sided DI-QKD on Alice's side as well as Bob's. For this demonstration, we set the source to a maximally entangled Bell state and measured in the 0°/90° ($\sigma_z$) and the 45°/135° ($\sigma_x$) bases, where $\sigma_i$ are the Pauli operators. In the $\sigma_z$-basis we observed a visibility of 99.78 % (± 0.03 %) and in the $\sigma_x$-basis a visibility of 96.78 % (± 0.01 %). Together with arm efficiencies 72.46 % (± 0.08 %) (78.12 % (± 0.09 %)) for Alice (Bob) and using the formulas[i] in (*30*), we estimate a minimum key rate of 1096 Hz (± 19 Hz) (461 Hz (± 11 Hz)) for one-sided DI-QKD on each side. An additional test with increased pump power yielded a four-fold increase in key rates, and we anticipate that some optimization in our setup and counting routines together with higher pump powers will yield significant further improvement on these values.

**Conclusion:** Using photons, we have demonstrated an experimental Bell inequality violation closing the fair-sampling loophole. Without relying on any assumed error distribution, we statistically verify a violation of the Eberhard inequality by nearly 70 standard deviations and thus clearly demonstrate the necessity to abandon all local realistic theories that take advantage of unfair sampling to explain the observed values. We note that this test makes photons the first system for which each of the three major loopholes has been closed, albeit in separate experiments. Moreover, the derivation of Eberhard's Bell inequality even includes events not detected on either side; hence no post-selection is necessary to violate the inequality. This is relevant not only to fundamental tests like a loophole-free Bell demonstration, but also represents promise for practical applications like, as demonstrated here, one-sided device-independent quantum key distribution, implemented from both sides.

**References and Notes:**

1. A. Einstein, B. Podolsky, N. Rosen, Can quantum-mechanical description of physical reality be considered complete? *Physical Review* **47**, 777–780 (1935).

2. J. S. Bell, On the Einstein Podolsky Rosen Paradox, *Physics* **1**, 195 (1964).

3. P. M. Pearle, Hidden-Variable Example Based upon Data Rejection, *Physical Review D* **2**, 1418–1425 (1970).

4. A. Garg, N. D. Mermin, Detector inefficiencies in the Einstein-Podolsky-Rosen experiment, *Physical Review D* **35**, 3831 (1987).

5. M. A. Rowe *et al.*, Experimental violation of a Bell's inequality with efficient detection, *Nature* **409**, 791–794 (2001).

6. M. Ansmann *et al.*, Violation of Bell's inequality in Josephson phase qubits, *Nature* **461**, 504–506 (2009).

7. D. N. Matsukevich, P. Maunz, D. L. Moehring, S. Olmschenk, C. Monroe, Bell Inequality Violation with Two Remote Atomic Qubits, *Physical Review Letters* **100**, 150404 (2008).

8. J. Hofmann *et al.*, Heralded Entanglement Between Widely Separated Atoms, *Science* **337**, 72–75 (2012).

9. S. J. Freedman, J. F. Clauser, Experimental test of local hidden-variable theories, *Physical Review Letters* **28**, 938–941 (1972).

10. A. Aspect, J. Dalibard, G. Roger, Experimental test of Bell's inequalities using time varying analyzers, *Physical Review Letters* **49**, 1804–1807 (1982).

11. G. Weihs, T. Jennewein, C. Simon, H. Weinfurter, A. Zeilinger, Violation of Bell's Inequality under Strict Einstein Locality Conditions, *Physical Review Letters* **81**, 5039–5043 (1998).

12. T. Scheidl *et al.*, Violation of local realism with freedom of choice, *Proceedings of the National Academy of Sciences* **107**, 19708–19713 (2010).

13. Z. Merali, Quantum Mechanics Braces for the Ultimate Test, *Science* **331**, 1380–1382 (2011).

14. P. H. Eberhard, Background Level and Counter Efficiencies Required for a Loophole-Free Einstein-Podolsky-Rosen Experiment, *Physical Review A* **47**, 747–750 (1993).

15. A. Fedrizzi, T. Herbst, A. Poppe, T. Jennewein, A. Zeilinger, A wavelength-tunable, fiber-coupled source of narrowband entangled photons, *Optics Express* **15**, 15377–15386 (2007).

16. S. Ramelow *et al.*, Highly efficient heralding of entangled single photons, *arXiv:12115059 [quant-ph]*.

17. A. E. Lita, A. J. Miller, S. W. Nam, Counting near-infrared single-photons with 95% efficiency, *Optics Express* **16**, 3032–3040 (2008).

18. J. F. Clauser, M. A. Horne, Experimental consequences of objective local theories, *Physical Review D* **10**, 526–535 (1974).

19. N. Brunner, Detection Loophole in Asymmetric Bell Experiments, *Physical Review Letters* **98**, 220403 (2007).

20. T. Vértesi, Closing the Detection Loophole in Bell Experiments Using Qudits, *Physical Review Letters* **104**, 060401 (2010).

21. J. F. Clauser, M. A. Horne, A. Shimony, R. A. Holt, Proposed Experiment to Test Local Hidden-Variable Theories, *Physical Review Letters* **23**, 880–884 (1969).

22. P. G. Kwiat, H. Eberhard, Philippe, A. M. Steinberg, R. Y. Chiao, Proposal for a loophole-free Bell inequality experiment, *Physical Review A* **49**, 3209–3220 (1994).

23. T. Kim, M. Fiorentino, F. N. C. Wong, Phase-stable source of polarization-entangled photons using a polarization Sagnac interferometer, *Physical Review A* **73**, 12316–12316 (2006).

24. W. P. Grice, R. S. Bennink, D. S. Goodman, A. T. Ryan, Spatial entanglement and optimal single-mode coupling, *Physical Review A* **83**, 023810 (2011).

25. D. Ljunggren, M. Tengner, Optimal focusing for maximal collection of entangled narrowband photon pairs into single-mode fibers, *Physical Review A* **72**, 062301 (2005).

26. D. Drung *et al.*, Highly Sensitive and Easy-to-Use SQUID Sensors, *IEEE Transactions on Applied Superconductivity* **17**, 699–704 (2007).

27. S. Pironio *et al.*, Device-independent quantum key distribution secure against collective attacks, *New Journal of Physics* **11**, 045021 (2009).

28. I. Gerhardt *et al.*, Experimentally Faking the Violation of Bell's Inequalities, *Physical Review Letters* **107**, 170404 (2011).

29. M. Lucamarini, G. Vallone, I. Gianani, P. Mataloni, G. Di Giuseppe, Device-independent entanglement-based Bennett 1992 protocol, *Physical Review A* **86**, 032325 (2012).

30. C. Branciard, E. Cavalcanti, S. Walborn, V. Scarani, One-sided device-independent quantum key distribution: Security, feasibility, and the connection with steering, *Physical Review A* **85**, 010301 (2012).

**Acknowledgments:** The authors acknowledge Marco Schmidt at the Physikalisch-Technische Bundesanstalt in Berlin, Germany for assistance with the TES-SQUID system setup.

This work was supported by the ERC (Advanced Grant QIT4QAD, 227844), the Austrian Science Foundation (FWF) under projects SFB F4008 and CoQuS, the grant Q-ESSENCE (no. 248095), QAP (No. 15848), the Marie Curie Research Training Network EMALI (No. MRTN-CT-2006-035369), and the John Templeton Foundation. This work was also supported by the NIST Quantum Information Science Initiative (QISI).

---

[i] For $q$ in (*30*), we assume $q=1$. Due to the quality of the components in our measurement apparatus, we estimate that $q$ will in fact be very close to 1.